\documentclass[letterpaper,12pt]{article}   %% LaTeX 2e (preferred)
\usepackage{osajnl2} %% do not use with REVTeX4
\usepackage[draft]{hyperref} %% optional
\usepackage{cite}
\usepackage{amsmath}
\usepackage{array} % for better arrays (eg matrices) in maths
\usepackage{astro_bib_macro}

\newcommand{\ro}{\pmb{\rho}}
\newcommand{\ko}{\pmb{K}}
\newcommand{\roprime}{\pmb{\ro^{\prime}}}
\newcommand{\tv}{\pmb{t}}

\begin{document}

\title{Transformation of Zernike coefficients:\\ A Fourier based method for scaled, translated and rotated wavefront apertures.}

\author{Eric Tatulli,$^{1,*}$}
\address{$^1$Inter-University Centre for Astronomy and Astrophysics, Ganeshkhind, Pune 411 007, India}
\address{$^*$Corresponding author: tatulli@iucaa.ernet.in}

\begin{abstract}This paper studies the effects on Zernike coefficients of aperture scaling, translation and rotation, when a given aberrated wavefront is described on the Zernike polynomial basis. It proposes a new analytical method for computing the matrix that enables the building of the transformed Zernike coefficients from the original ones. The technique is based on the properties of Zernike polynomials Fourier Transform and, in the case of a full aperture without central obstruction, the coefficients of the matrix are given in terms of integrals of Bessel functions. The integral formulas are exact and do not depend on any specific ordering of the polynomials.  
\end{abstract}

\ocis{070.0070, 000.3860, 350.1260, 010.1290, 330.4460.}

\maketitle 

\section{Introduction}
The use of Zernike polynomials for representing the aberrations of an optical system is now common and well understood \cite{born_1}. Indeed, these polynomials have been proven to form a very convenient basis in  applications as diverse as the description of pupil aberrations of the human eye \cite{thibos_1} or the characterization  in observational astronomy of the statistics of turbulent aberrations produced by the atmosphere over the telescope aperture \cite{fried_1}. When using the Zernike basis, the aberrated phase is described as a linear combination of the polynomials, the value of the coefficients being related to the aperture on which the phase is defined. When this aperture undergoes geometrical transformations such as scaling, translation and rotation, the Zernike coefficients corresponding to the transformed aperture can be computed from proper linear combinations of the original ones by finding the transformation matrix associated to these linear applications.\\
In that matter, if many solutions have been proposed for pure pupil scaling, see e.g. \cite{goldberg_1,schwiegerling_1,campbell_1,shu_1,janssen_1}, only a few authors have tackled the three linear transformations altogether in a unified method. Let us mention Lundstr{\"o}m \& Unsbo \cite{lundstrom_1} who, by generalizing the paper of Campbell \cite{campbell_1}, have developed an ingenious technique that however necessitates a specific reordering of the polynomials in the complex plane. Alternatively Bara \textit{et al.} \cite{bara_1} have proposed a method based on the theoretically simple principle of reference change matrix built from a given grid of points that sample the wavefront. In practice however this technique appears sometimes difficult to implement, especially when the scaling factor is important and/or when the number of Zernike coefficients needed to represent the phase is large (the matrix being in these cases often ill-conditioned).\\
In this paper we propose an alternative technique based on the Fourier Transform properties of the Zernike polynomials. We provide analytical and exact formulas that allow the computation of the transformation matrix in all cases of scaling/translation/rotation of the pupil. In the hypothesis of a full aperture without central obstruction, we develop the results in terms of integrals of a product of two or three Bessel functions. The numerical evaluation of these integrals can be fastly performed by making use of Mellin Transform properties \cite{sasiela_1} or merely by referring to published tables of integrals \cite{luke_1, gradshteyn_1} and it will not be addressed here. Section (\ref{sec_zernprop}) recalls the properties of the Zernike polynomials useful for the formal derivation of our calculations. Section (\ref{sec_changebase})provides the general formalism that enablesthe building of the transformation matrix, whereas Sects (\ref{sec_scale}, \ref{sec_trans}, \ref{sec_rot}), are focusing in further details on the scaling, translation and rotation processes, respectively.

\section{Zernike polynomials basic properties}\label{sec_zernprop}
Any  wavefront  $\phi(\pmb{r})$ defined over a circular aperture can be described as a linear combination of Zernike polynomials  $Z_j(\ro)$:
\begin{equation}
\phi(\pmb{r}) = \phi(R\ro) = \sum^{\infty}_{j=1} a_j Z_j(\ro)
\end{equation}
where $R$ is the physical radius of the pupil. Zernike polynomials are defined over the circle of unit radius $\Pi_p(\ro)$ , that is: 
\begin{eqnarray}
\Pi_p(\ro) = \left\{\begin{array}{cl} 1/\pi & \rm{if}~|\ro| \le 1 \\ 0 & \rm{elsewhere}\end{array}\right. \label{unitpupil}
\end{eqnarray}
The Zernike coefficients can be calculated by projecting the phase on the Zernike  basis: 
\begin{equation}
a_j  = \int  \Pi_p(\ro) Z_j(\ro)  \phi(R\ro) \mathrm{d}^2\ro \label{coeff_ai}
\end{equation}
In polar coordinates, Zernike polynomials are defined for a circular aperture without obstruction as:
\begin{equation}
Z^m_n(\rho, \theta) = Z_j(\rho, \theta) = \sqrt{n+1}R^m_n(\rho) \left\{\begin{array}{l}\sqrt{2}\cos(|m|\theta)~~\mathrm{if}~m>0\\ \sqrt{2}\sin(|m|\theta)~~\mathrm{if}~m<0\\1~~\mathrm{if}~m=0\end{array}\right.
\end{equation}
where $n$ and $m$  are respectively the radial degree and the azimuthal frequency of the $j^{th}$ polynomial (note that $n$ and  $m$ have necessarily the same parity), $j$ being defined as $j=\frac{n(n+2)+m}{2}+1$, and:
 \begin{equation}
R^m_n(\rho)  = \sum_{s=0}^{(n-|m|)/2} \frac{(-1)^s(n-s)!}{s![(n+|m|)/2-s]![(n-|m|)/2-s]!}\rho^{n-2s}
\end{equation}
In the full aperture case, Zernike polynomials are orthonormal:
\begin{equation}
\int \Pi_p(\ro)Z_i(\ro)Z_j(\ro) \mathrm{d}^2\ro = \delta_{ij}
\end{equation}
and the Fourier Transform $Q_j(\kappa, \alpha)$ of $\Pi_p(\ro)Z_j(\ro)$ writes in polar coordinates:
\begin{equation}
Q_j(\kappa, \alpha) = (-1)^n\sqrt{n+1}\frac{J_{n+1}(2\pi\kappa)}{\pi\kappa}\left\{\begin{array}{l}(-1)^{(n-|m|)/2}i^{|m|}\sqrt{2}\cos(|m|\alpha)~~\mathrm{if}~m>0\\ (-1)^{(n-|m|)/2}i^{|m|}\sqrt{2}\sin(|m|\alpha)~~\mathrm{if}~m<0\\(-1)^{n/2}~~\mathrm{if}~m=0\end{array}\right. \label{eq_zern_fourier}
\end{equation}
\section{Changing axis reference: building the transformation matrix} \label{sec_changebase}
\begin{figure}[t]
\begin{center}
\includegraphics[width=0.5\textwidth]{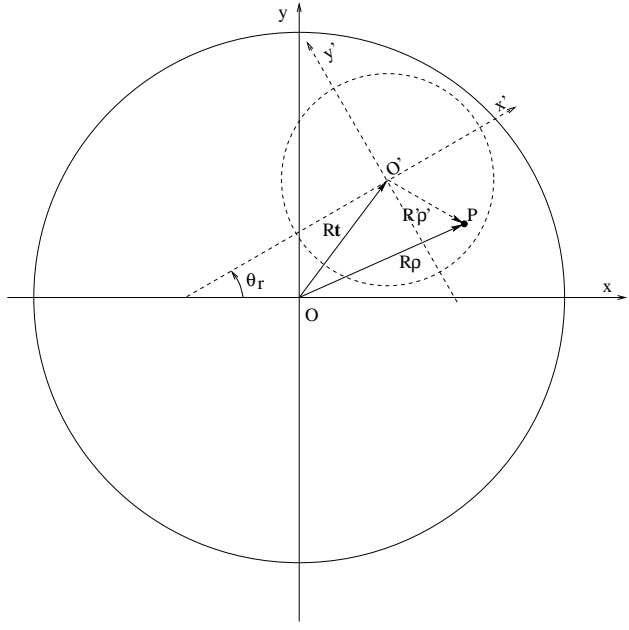}
\caption{\label{fig_coord} Sketch of the reference systems used in Sect. (\ref{sec_changebase}). In the $\widehat{Oxy}$ reference axis (solid lines), the phase is defined over the aperture of radius $R$ and the unitary coordinates of a point $P$ in this system is $\ro$. In the $\widehat{O^{\prime}x^{\prime}y^{\prime}}$ reference axis (dashed lines) the phase is defined over the aperture of radius $R^{\prime}$ and the unitary coordinates of a point $P$ in this system is $\roprime$. The reference change involves a translation of vector $R\tv$ and a rotation of angle $\theta_r$.}
\end{center}
\end{figure}Figure (\ref{fig_coord}) sketches the principle of reference system shifting. The input phase $\phi(R\ro)$ is defined over the pupil diameter $R$, in the $\widehat{Oxy}$ reference axis. The output phase $\phi(R^{\prime}\roprime)$ defined over the pupil diameter $R^{\prime}$ is a scaled/translated/rotated version of the input one, and is centered on the $\widehat{O^{\prime}x^{\prime}y^{\prime}}$ reference axis.  We call $\tv$ the unitary translation vector between both reference systems, and $\theta_r$ the rotation angle. For a given point $P$ of the phase that belongs to both input and output pupils, we have $\phi(P) = \phi(R\ro) =  \phi(R^{\prime}\roprime)$ where:
\begin{equation}
R^{\prime}\roprime= \mathcal{R}_{\theta_r}\left[R\ro - R\tv\right] \Leftrightarrow \roprime = \mathcal{R}_{\theta_r}\left[\beta(\ro-\tv)\right]   \label{eq_vectors}
\end{equation}    
with $\beta = R/R^{\prime}$ and $\mathcal{R}_{\theta_r}$ is the rotation operator. In the following we impose the transformed pupil to lie inside the input one (no extrapolation), that is $R$, $R^{\prime}$ and $\tv$ have to verify this additional relationship:
\begin{equation}
R|\tv| + R^{\prime} \le R  \Leftrightarrow \beta|\tv|+1 \le \beta \label{eq_condition}
\end{equation}
In this framework the input phase is described as a sum of $N_z$ Zernike polynomials:
\begin{equation}
\phi(R\ro) = \sum^{N_z}_{j=1} a_j Z_j(\ro)
\end{equation}
Ragazzoni \textit{et al.} \cite{ragazzoni_1} and Shu \textit{et al.} \cite{shu_1} have demonstrated that it exists a set of Zernike coefficients limited by the same highest polynomial number $N_z$ that can represent the wavefront over the transformed pupil, that is:
\begin{equation}
\phi(R^{\prime}\roprime) = \sum^{N_z}_{i=1} b_i Z_i(\roprime)
\end{equation}
with
\begin{equation}
b_i  = \int  \Pi_p(\roprime) Z_i(\roprime)  \phi(R^{\prime}\roprime) \mathrm{d}^2{\roprime} \label{coeff_bi}
\end{equation}
Inside the output pupil, we have $\phi(R^{\prime}\roprime) = \phi(R\ro)$ and we can modify previous equation as:
\begin{equation}
b_i  = \int  \Pi_p(\roprime) Z_i(\roprime)  \phi(R\ro) \mathrm{d}^2{\roprime} = \sum^{N_z}_{j=1} a_j \int  \Pi_p(\roprime) Z_i(\roprime) Z_j(\ro) \mathrm{d}^2{\roprime} 
\end{equation}
that we can rewrite:
\begin{equation}
b_i  = \pi \sum^{N_z}_{j=1} a_j \int  \Pi_p(\roprime) Z_i(\roprime)  \Pi_p(\ro) Z_j(\ro) \mathrm{d}^2{\roprime} 
\end{equation}
providing that Eq. (\ref{eq_condition}) is verified. Then using Eq. (\ref{eq_vectors}) it comes:
\begin{equation}
b_i  =  \pi \sum^{N_z}_{j=1} a_j \int  \Pi_p(\roprime) Z_i(\roprime) \Pi_p\left(\frac{\mathcal{R}_{-\theta_r}[\roprime]}{\beta}+\tv\right) Z_j\left(\frac{\mathcal{R}_{-\theta_r}[\roprime]}{\beta}+\tv\right) \mathrm{d}^2{\roprime} 
\end{equation}
Making use of convolution and Fourier Transform properties \cite{molodij_1}, the latter equation writes:
\begin{equation}
b_i  =  \beta^2 \pi  \sum^{N_z}_{j=1} a_j \int Q^{\ast}_i(\ko) Q^{\mathcal{R}_{-\theta_r}}_j(\beta\ko) \exp\left[2i\pi\beta\tv.\ko\right] {d}^2{\ko} \label{eq_bi}
\end{equation}
where $Q^{\mathcal{R}_{\theta_r}}_j(\ko)$ denotes the Fourier Transform of $\Pi_pZ_j\left(\mathcal{R}_{\theta_r}[\ro]\right)$. Note that in polar coordinates $\Pi_pZ_j\left(\mathcal{R}_{\theta_r}[\ro]\right) = \Pi_pZ_j(\rho, \theta-\theta_r)$, hence its Fourier Transform simply translates as $Q^{\mathcal{R}_{\theta_r}}_j(\kappa, \alpha)=Q_j(\kappa, \alpha-\theta_r)$. Rewriting Eq. (\ref{eq_bi}) as a product of matrices, we finally have:
\begin{equation}
\pmb{b} = M^{[\beta,t,\theta_r]}.\pmb{a} 
\end{equation}
where  $\pmb{a} = [a_1,...,a_{N_z}]$, $\pmb{b} = [b_1,...,b_{N_z}]$, and the coefficients $(M^{[\beta,t,\theta_r]}_{ij})$ of the matrix $M^{[\beta,t,\theta_r]}$ verify:
\begin{equation}
M^{[\beta,t,\theta_r]}_{ij} = \pi \beta^2  \int_0^{\infty}\int_0^{2\pi}  \kappa Q^{\ast}_i(\kappa, \alpha) Q_j(\beta\kappa, \alpha+\theta_r) \exp\left[2i\pi\beta|\tv|\kappa \cos(\alpha-\alpha_t)\right]  {d}{\alpha} {d}{\kappa} \label{eq_tmatrix}
\end{equation}
with $\alpha_t = \mathrm{arg}(\tv)$. Eq. (\ref{eq_tmatrix}) is the general expression of the linear application that enables the transformation of the Zernike coefficients in cases of scaling/translation/rotation of the pupil over which the wavefront is originally defined. In the following, we will focus on the full aperture case without central obstruction such that Eq. (\ref{eq_zern_fourier}) can be used for our further derivations.

\section{Concentric pupils}\label{sec_scale}
\begin{figure*}[!h]
\begin{center}
\begin{tabular}{c}
\includegraphics[width=\textwidth]{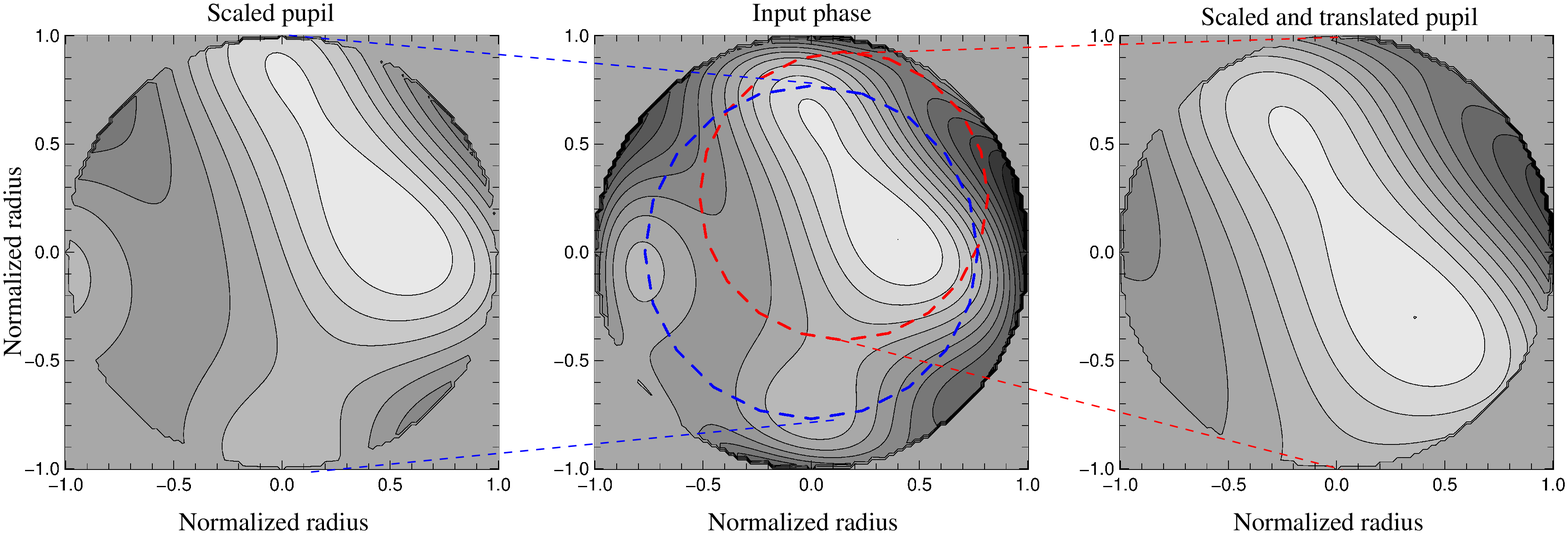}\\\includegraphics[width=\textwidth]{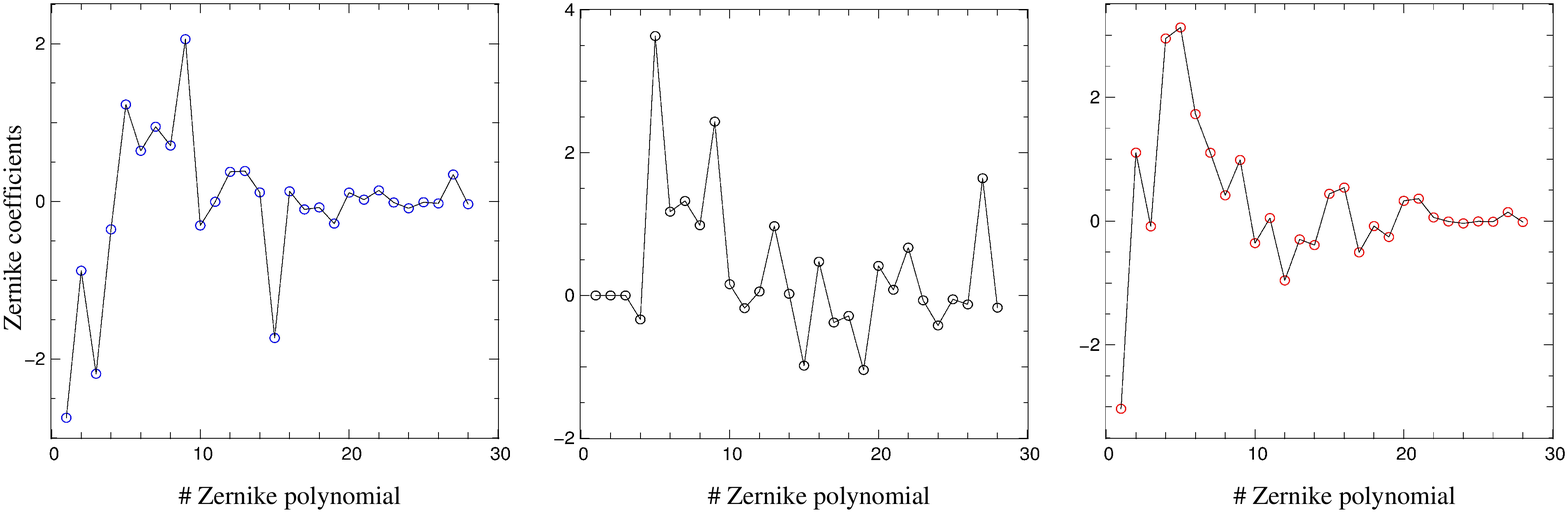}
\end{tabular}
\caption{\label{fig_trans_wavefront} Center: initial phase (top) and associated Zernike coefficients (bottom), with $j=1$ to $28$, that is the phase is defined with $n=6$ radial degree modes. Left: reconstructed phase over the scaled aperture ($\beta=1.3$) computed from the transformed Zernike coefficients. These coefficients are calculated from the original ones using both the transformation matrix of Eq. (\ref{eq_scaling_final}) (circles) and Shu \textit{et al.} method \cite{shu_1} (solid line) . Right: same as previously for a scaled and translated aperture ($\beta=1.5$, $|\tv|=0.3$, $\alpha_t=\pi/3$). The transformed coefficients derived from Eq. (\ref{eq_scalingtrans_final}) (circles) are compared with that of Lundstr{\"o}m \& Unsbo technique \cite{lundstrom_1} (solid line).}
\end{center}
\end{figure*}
We first investigate the case where the output pupil is a shrunk version of the input one with a scaling factor of $R^{\prime}/R = 1/\beta$ (and with $|\tv|=0$, $\alpha_t=0$, $\theta_r=0$). Introducing the expression of Zernike Fourier Transform in Eq. (\ref{eq_tmatrix}), it comes:
\begin{eqnarray}
M^{[\beta]}_{ij} &=&  \frac{\beta}{\pi} (-1)^{\frac{3(n_i+n_j)}{2}}(-1)^{\frac{-(|m_i|+|m_j|)}{2}}i^{(|m_j|-|m_i|)}\sqrt{n_i+1}\sqrt{n_j+1}\nonumber \\
&& \times  \int_0^{\infty} \kappa^{-1}J_{n_i+1}(2\pi\kappa)J_{n_j+1}(2\pi\beta\kappa) {d}{\kappa} \nonumber \\ 
&& \times \int_0^{2\pi} \left[\begin{array}{l} \sqrt{2}\cos(|m_i|\alpha)\\ \sqrt{2}\sin(|m_i|\alpha)\\1\end{array}\right] \bigotimes \left[\begin{array}{l} \sqrt{2}\cos(|m_j|\alpha)\\ \sqrt{2}\sin(|m_j|\alpha)\\1\end{array}\right]{d}{\alpha} \label{eq_scaling}
\end{eqnarray}
where the operator $\bigotimes$ denotes all the possible multiplicative combinations between the coefficients of the vectors, depending on the respective values of $m_i$ and $m_j$ (i.e. 9 multiplications possible in this case). From the trigonometric properties of cosine functions, it comes straightforward that the integral over $\alpha$ in Eq. (\ref{eq_scaling}) is non zero only if $m_i = m_j$. Hence previous equation can be replaced by: 
\begin{equation}
M^{[\beta]}_{ij} = \left\{\begin{array}{l} 2\beta\sqrt{n_i+1}\sqrt{n_j+1} \int_0^{\infty} \kappa^{-1}J_{n_i+1}(2\pi\kappa)J_{n_j+1}(2\pi\beta\kappa) {d}{\kappa}~~\mathrm{if}~m_i=m_j\\0 ~~\mathrm{if}~m_i \neq m_j \end{array}\right. \label{eq_scaling_final}
\end{equation}
Figure (\ref{fig_trans_wavefront}, left and center) shows an example of pupil scaling with $\beta=1.3$. We can verify on the upper plots  that the computed phase over the scaled pupil is indeed the zoomed version of the input one, with the appropriate zooming factor $\beta$. Bottom graphics are displaying the input and derived output Zernike coefficients. In order to further check the validity of our method, we have also plotted the estimated output coefficients calculated with that of Shu \textit{et al.} \cite{shu_1}. Both independent techniques are giving the same results.

\section{Scaling and translating}\label{sec_trans}
Since we have assumed that the transformed pupil should lie within the original one, any shifting must necessarily be combined with appropriate scaling, such that Eq. (\ref{eq_condition}) is verified. As a consequence, pure translation process cannot be tackled by this study unless a so-called \textit{meta-}pupil encompassing both pre- and post-shifted apertures is considered beforehand \cite{ragazzoni_1}. Providing that this hypothesis is verified, Eq. (\ref{eq_tmatrix}) rewrites (with $\theta_r=0$):
\begin{eqnarray}
&&M^{[\beta,t]}_{ij} = \frac{\beta}{\pi} (-1)^{\frac{3(n_i+n_j)-|m_i|-|m_j|}{2}}i^{(|m_j|-|m_i|)}\sqrt{n_i+1}\sqrt{n_j+1}\int_0^{\infty} \kappa^{-1}J_{n_i+1}(2\pi\kappa)J_{n_j+1}(2\pi\beta\kappa)\nonumber \\
&&\times \int_0^{2\pi} \left[\begin{array}{l} \sqrt{2}\cos(|m_i|\alpha)\\ \sqrt{2}\sin(|m_i|\alpha)\\1\end{array}\right] \bigotimes \left[\begin{array}{l} \sqrt{2}\cos(|m_j|\alpha)\\ \sqrt{2}\sin(|m_j|\alpha)\\1\end{array}\right] \exp\left[2i\pi\beta|\tv|\kappa \cos(\alpha-\alpha_t)\right]{d}{\alpha} {d}{\kappa}  \label{eq_scalingtrans}
\end{eqnarray}
The derivation of the integral over $\alpha$ requires to make use of the integral definition of Bessel functions \cite{watson_1}, that is:
\begin{equation}
J_n(\kappa) = \frac{1}{2\pi}\int_{-\pi}^{\pi} \exp \left[-i(n\alpha-\kappa\sin\alpha)\right] {d}{\alpha}~~\mathrm{(with}~n~\mathrm{integer)}
\end{equation}
The formal development of Eq. (\ref{eq_scalingtrans}) necessitates several steps that are not difficult however somewhat tedious. They are presented in App. (A). The coefficients of the matrix $M^{[\beta,t]}$ eventually write:
\begin{eqnarray}
&M^{[\beta,t]}_{ij} =& \sqrt{2}^{\delta_{m_i,0}.\delta_{m_j,0}}(-1)^{\frac{3(n_i+n_j)-|m_i|-|m_j|}{2}}i^{(|m_j|-|m_i|)}\beta\sqrt{n_i+1}\sqrt{n_j+1} \nonumber \\
&&\times \left[\mathcal{A}^{-}_{ij,\alpha_t}\int_0^{\infty} \kappa^{-1}J_{n_i+1}(2\pi\kappa)J_{n_j+1}(2\pi\beta\kappa)J_{||m_i|-|m_j||}(2\pi\beta|\tv|\kappa){d}{\kappa} \right. \nonumber  \\
&&\hskip20pt   + \mathcal{A}^{+}_{ij,\alpha_t}\left.\int_0^{\infty} \kappa^{-1}J_{n_i+1}(2\pi\kappa)J_{n_j+1}(2\pi\beta\kappa)J_{|m_i|+|m_j|}(2\pi\beta|\tv|\kappa)]{d}{\kappa}\right] \label{eq_scalingtrans_final}
\end{eqnarray}
with
\begin{eqnarray}
\mathcal{A}^{-}_{ij,\alpha_t} = (-1)^{\frac{||m_i|-|m_j||}{2}} \left\{\begin{array}{ll} \cos([|m_i|-|m_j|]\alpha_t)&\mathrm{if}~\mathrm{sgn}(m_i)=\mathrm{sgn}(m_j)\\\mathrm{sgn}(m_i)\sin([|m_i|-|m_j|]\alpha_t)&\mathrm{if}~\mathrm{sgn}(m_i)\neq\mathrm{sgn}(m_j) \end{array}\right. \nonumber
\end{eqnarray}
and
\begin{eqnarray}
\mathcal{A}^{+}_{ij,\alpha_t} = (-1)^{\frac{||m_i|+|m_j||}{2}} \left\{\begin{array}{ll}\mathrm{sgn}(m_i)\cos([|m_i|+|m_j|]\alpha_t)&\mathrm{if}~\mathrm{sgn}(m_i)=\mathrm{sgn}(m_j)\\\sin([|m_i|+|m_j|]\alpha_t)&\mathrm{if}~\mathrm{sgn}(m_i)\neq\mathrm{sgn}(m_j) \end{array}\right. \nonumber
\end{eqnarray}
If Eq. (\ref{eq_scalingtrans_final}) appears to return complex numbers, the conjugated parity of $n_i$ and $m_i$ (respectively $n_j$ and $m_j$) insures to produce a real coefficient for any pair of Zernike polynomials considered. Figure (\ref{fig_trans_wavefront}, center and right) illustrates the case of scaling + translation with $\beta=1.5$, $|\tv|=0.3$, and $\theta_t=\pi/3$ and results of the present method are proven to match with that of Lundstr{\"o}m \& Unsbo \cite{lundstrom_1}.
\section{Rotation}\label{sec_rot}
\begin{figure*}[t]
\begin{center}
\includegraphics[width=\textwidth]{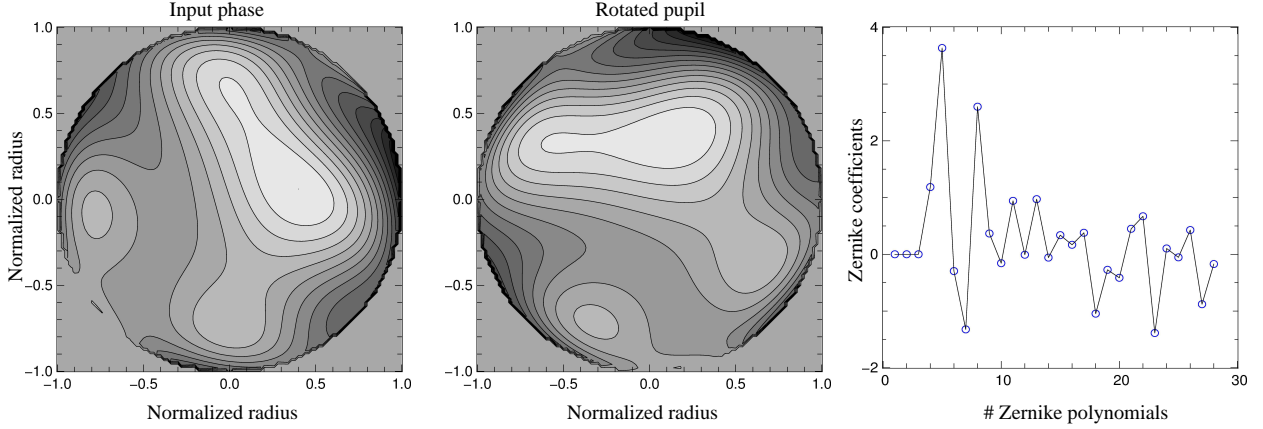}
\caption{\label{fig_rotation_wavefront} Left: original wavefront. Center: reconstructed wavefront for a pupil rotation of $\theta_r=\pi/3$. Right: transformed Zernike coefficients derived from the original ones using Eq. (\ref{eq_rotation_final_final}) (circles) and Bara \textit{et al.}  method \cite{bara_1} (solid lines).}
\end{center}
\end{figure*}
We finally investigate the case of pure rotation as it can be treated independently from the scaling and translation processes. A full scaling+translation+rotation transformation can indeed be decomposed in two consecutive processes, that is scaling+translation then pure rotation (or similarly rotation first, then scaling+translation). In other words the transformation matrix of the full transformation $M^{[\beta,t,\theta_r]}$ can be computed as:
\begin{equation}
M^{[\beta,t,\theta_r]} = M^{[\beta,t]}.M^{[\theta_r]} = M^{[\theta_r]}.M^{[\beta,t]}
\end{equation} 
Applying Eq. (\ref{eq_tmatrix}) with $\beta=1$ and $|\tv|=0$, it comes:
\begin{eqnarray}
M^{[\theta_r]}_{ij} &=&  \frac{1}{\pi} (-1)^{\frac{3(n_i+n_j)}{2}}(-1)^{\frac{-(|m_i|+|m_j|)}{2}}i^{(|m_j|-|m_i|)}\sqrt{n_i+1}\sqrt{n_j+1}\nonumber \\
&& \times  \int_0^{\infty} \kappa^{-1}J_{n_i+1}(2\pi\kappa)J_{n_j+1}(2\pi\kappa) {d}{\kappa} \nonumber \\ 
&& \times \int_0^{2\pi} \left[\begin{array}{l} \sqrt{2}\cos(|m_i|\alpha)\\ \sqrt{2}\sin(|m_i|\alpha)\\1\end{array}\right] \bigotimes \left[\begin{array}{l} \sqrt{2}\cos(|m_j|[\alpha+\theta_r])\\ \sqrt{2}\sin(|m_j|[\alpha+\theta_r])\\1\end{array}\right]{d}{\alpha} \label{eq_rotation}
\end{eqnarray}
In this case, trigonometric properties of cosine functions imply that the integral over $\alpha$ is non-zero only when $|m_i| = |m_j|$, and Eq. (\ref{eq_rotation}) reduces to:
\begin{eqnarray}
M^{[\theta_r]}_{ij} &=&  2\sqrt{n_i+1}\sqrt{n_j+1} \int_0^{\infty} \kappa^{-1}J_{n_i+1}(2\pi\kappa)J_{n_j+1}(2\pi\kappa) {d}{\kappa} \nonumber \\
&&\times\left\{\begin{array}{l}\cos(m_j\theta_r)~~\mathrm{if}~m_i=m_j\\\sin(m_j\theta_r)~~\mathrm{if}~m_i=-m_j~~\mathrm{(and}~m_i \neq 0\mathrm{)}\\0 ~~\mathrm{if}~|m_i| \neq |m_j| \end{array}\right. \label{eq_rotation_final}
\end{eqnarray}
Furthermore it can be shown that:
\begin{equation}
2\sqrt{n_i+1}\sqrt{n_j+1} \int_0^{\infty} \kappa^{-1}J_{n_i+1}(2\pi\kappa)J_{n_j+1}(2\pi\kappa) {d}{\kappa} = \delta_{n_i,n_j}
\end{equation}
so that $M^{[\theta_r]}_{ij}$ takes the simpler form:
\begin{equation}
M^{[\theta_r]}_{ij} = \delta_{n_i,n_j}\times\left\{\begin{array}{l}\cos(m_j\theta_r)~~\mathrm{if}~m_i=m_j\\\sin(m_j\theta_r)~~\mathrm{if}~m_i=-m_j~~\mathrm{(and}~m_i \neq 0\mathrm{)}\\0 ~~\mathrm{if}~|m_i| \neq |m_j| \end{array}\right. \label{eq_rotation_final_final}
\end{equation}
As a consequence, the rotation matrix does not necessitate the evaluation of integral of Bessel functions but merely requires to compute cosine functions for multiples of the rotation angle accordingly to the accounted azimuthal frequencies ($m_j$). Figure (\ref{eq_rotation_final_final}) shows an example of phase rotation with $\theta_r=\pi/3$. The transformed Zernike coefficients calculated from Eq. (\ref{eq_rotation_final_final}) are successfully compared with that of Bara \textit{et al.} \cite{bara_1}.
\section{Summary}
This paper provides analytical formulas that enable the computation of Zernike coefficients for scaling, translation and rotation of the aperture over which the wavefront is originally defined. The expressions of the transformation matrix that allows to switch from the original Zernike coefficients to the transformed ones are exact. Each coefficient of the matrix can be calculated directly and independently without involving any specific ordering of the polynomials. Our method has been compared to several independent techniques avalaible in the literature. The results are consistently matching, hence validating the approach presented in this paper.

\appendix
\renewcommand{\theequation}{A-\arabic{equation}}
\setcounter{equation}{0} 

\section*{Appendix A: Computing the scaling+translation matrix} \label{app_scalingtrans}
We focus here on the integral over $\alpha$ of Eq. (\ref{eq_scalingtrans}):
\begin{equation}
F(\kappa) = \int_0^{2\pi} \left[\begin{array}{l} \sqrt{2}\cos(|m_i|\alpha)\\ \sqrt{2}\sin(|m_i|\alpha)\\1\end{array}\right] \bigotimes \left[\begin{array}{l} \sqrt{2}\cos(|m_j|\alpha)\\ \sqrt{2}\sin(|m_j|\alpha)\\1\end{array}\right] \exp\left[2i\pi\beta|\tv|\kappa \cos(\alpha-\alpha_t)\right]{d}{\alpha}   \label{eq_scalingtrans_alpha}
\end{equation}
We recall the integral definitions of Bessel functions that will be used to unfold Eq. (\ref{eq_scalingtrans_alpha}):
\begin{eqnarray}
&&\int_0^{2\pi} \cos(m\alpha)\exp(iy\cos(\alpha-\alpha_t)) \rm{d}\alpha = \left\{\begin{array}{l}2\pi(-1)^\frac{|m|}{2}\cos(m\alpha_t)J_{|m|}(y)~~ \mathrm{if~}m~\mathrm{even} \\ 2i\pi(-1)^\frac{|m|-1}{2}\cos(m\alpha_t)J_{|m|}(y)~~ \mathrm{if~}m~\mathrm{odd}\end{array}\right. \label{eq_jm_cos}\\
&&\int_0^{2\pi} \sin(m\alpha)\exp(iy\cos(\alpha-\alpha_t)) \rm{d}\alpha = \left\{\begin{array}{l}2\pi(-1)^\frac{|m|}{2}\sin(m\alpha_t)J_{|m|}(y)~~ \mathrm{if~}m~\mathrm{even} \\ 2i\pi(-1)^\frac{|m|-1}{2}\sin(m\alpha_t)J_{|m|}(y)~~ \mathrm{if~}m~\mathrm{odd}\end{array}\right. \label{eq_jm_sin}
\end{eqnarray}
We now have to investigate the different combinations regarding $m_i$ and $m_j$.
\begin{enumerate}
\item \underline{$m_i>0$ and $m_j>0$:} Eq. (\ref{eq_scalingtrans_alpha}) rewrites:
\begin{eqnarray}
F(\kappa) &=& \int_0^{2\pi} \cos([|m_i|-|m_j|]\alpha) \exp\left[2i\pi\beta|\tv|\kappa \cos(\alpha-\alpha_t)\right]{d}{\alpha}\nonumber \\
&+& \int_0^{2\pi} \cos([|m_i|+|m_j|]\alpha)\exp\left[2i\pi\beta|\tv|\kappa \cos(\alpha-\alpha_t)\right]{d}{\alpha}\nonumber
\end{eqnarray}
that, according to Eq. (\ref{eq_jm_cos}) gives:
\begin{enumerate}
\item if $m_i$ and $m_j$ have same parity:
\begin{eqnarray}
F(\kappa) &=& 2\pi (-1)^\frac{||m_i|-|m_j||}{2}\cos([|m_i|-|m_j|]\alpha_t) J_{||m_i|-|m_j||}(2\pi\beta|\tv|\kappa)\nonumber \\
&+& 2\pi (-1)^\frac{||m_i|+|m_j||}{2}\cos([|m_i|+|m_j|]\alpha_t) J_{||m_i|+|m_j||}(2\pi\beta|\tv|\kappa)
\end{eqnarray}
\item if $m_i$ and $m_j$ have different parity:
\end{enumerate}
\begin{eqnarray}
F(\kappa) &=& 2i\pi (-1)^\frac{||m_i|-|m_j||-1}{2}\cos([|m_i|-|m_j|]\alpha_t) J_{||m_i|-|m_j||}(2\pi\beta|\tv|\kappa)\nonumber \\
&+& 2i\pi (-1)^\frac{||m_i|+|m_j||-1}{2}\cos([|m_i|+|m_j|]\alpha_t) J_{||m_i|+|m_j||}(2\pi\beta|\tv|\kappa)
\end{eqnarray}
\item \underline{$m_i<0$ and $m_j<0$:}
\begin{eqnarray}
F(\kappa) &=& \int_0^{2\pi} \cos([|m_i|-|m_j|]\alpha) \exp\left[2i\pi\beta|\tv|\kappa \cos(\alpha-\alpha_t)\right]{d}{\alpha}\nonumber \\
&-& \int_0^{2\pi} \cos([|m_i|+|m_j|]\alpha)\exp\left[2i\pi\beta|\tv|\kappa \cos(\alpha-\alpha_t)\right]{d}{\alpha}\nonumber
\end{eqnarray}
that rewrites:
\begin{enumerate}
\item if $m_i$ and $m_j$ have same parity:
\begin{eqnarray}
F(\kappa) &=& 2\pi (-1)^\frac{||m_i|-|m_j||}{2}\cos([|m_i|-|m_j|]\alpha_t) J_{||m_i|-|m_j||}(2\pi\beta|\tv|\kappa)\nonumber \\
&-& 2\pi (-1)^\frac{||m_i|+|m_j||}{2}\cos([|m_i|+|m_j|]\alpha_t) J_{||m_i|+|m_j||}(2\pi\beta|\tv|\kappa)
\end{eqnarray}
\item if $m_i$ and $m_j$ have different parity:
\end{enumerate}
\begin{eqnarray}
F(\kappa) &=& 2i\pi (-1)^\frac{||m_i|-|m_j||-1}{2}\cos([|m_i|-|m_j|]\alpha_t) J_{||m_i|-|m_j||}(2\pi\beta|\tv|\kappa)\nonumber \\
&-& 2i\pi (-1)^\frac{||m_i|+|m_j||-1}{2}\cos([|m_i|+|m_j|]\alpha_t) J_{||m_i|+|m_j||}(2\pi\beta|\tv|\kappa)
\end{eqnarray}
\item \underline{$m_i<0$ and $m_j>0$:}
\begin{eqnarray}
F(\kappa) &=& - \int_0^{2\pi} \sin([|m_i|-|m_j|]\alpha) \exp\left[2i\pi\beta|\tv|\kappa \cos(\alpha-\alpha_t)\right]{d}{\alpha}\nonumber \\
&+& \int_0^{2\pi} \sin([|m_i|+|m_j|]\alpha)\exp\left[2i\pi\beta|\tv|\kappa \cos(\alpha-\alpha_t)\right]{d}{\alpha}\nonumber
\end{eqnarray}
that, according to Eq. (\ref{eq_jm_sin}) gives:
\begin{enumerate}
\item if $m_i$ and $m_j$ have same parity:
\begin{eqnarray}
F(\kappa) &=& -2\pi (-1)^\frac{||m_i|-|m_j||}{2}\sin([|m_i|-|m_j|]\alpha_t) J_{||m_i|-|m_j||}(2\pi\beta|\tv|\kappa)\nonumber \\
&+& 2\pi (-1)^\frac{||m_i|+|m_j||}{2}\sin([|m_i|+|m_j|]\alpha_t) J_{||m_i|+|m_j||}(2\pi\beta|\tv|\kappa)
\end{eqnarray}
\item if $m_i$ and $m_j$ have different parity:
\end{enumerate}
\begin{eqnarray}
F(\kappa) &=& -2i\pi (-1)^\frac{||m_i|-|m_j||-1}{2}\sin([|m_i|-|m_j|]\alpha_t) J_{||m_i|-|m_j||}(2\pi\beta|\tv|\kappa)\nonumber \\
&+& 2i\pi (-1)^\frac{||m_i|+|m_j||-1}{2}\sin([|m_i|+|m_j|]\alpha_t) J_{||m_i|+|m_j||}(2\pi\beta|\tv|\kappa)
\end{eqnarray}
\item \underline{$m_i>0$ and $m_j<0$:}
\begin{eqnarray}
F(\kappa) &=& \int_0^{2\pi} \sin([|m_i|-|m_j|]\alpha) \exp\left[2i\pi\beta|\tv|\kappa \cos(\alpha-\alpha_t)\right]{d}{\alpha}\nonumber \\
&+& \int_0^{2\pi} \sin([|m_i|+|m_j|]\alpha)\exp\left[2i\pi\beta|\tv|\kappa \cos(\alpha-\alpha_t)\right]{d}{\alpha}\nonumber
\end{eqnarray}
\begin{enumerate}
\item if $m_i$ and $m_j$ have same parity:
\begin{eqnarray}
F(\kappa) &=& 2\pi (-1)^\frac{||m_i|-|m_j||}{2}\sin([|m_i|-|m_j|]\alpha_t) J_{||m_i|-|m_j||}(2\pi\beta|\tv|\kappa)\nonumber \\
&+& 2\pi (-1)^\frac{||m_i|+|m_j||}{2}\sin([|m_i|+|m_j|]\alpha_t) J_{||m_i|+|m_j||}(2\pi\beta|\tv|\kappa)
\end{eqnarray}
\item if $m_i$ and $m_j$ have different parity:
\end{enumerate}
\begin{eqnarray}
F(\kappa) &=& 2i\pi (-1)^\frac{||m_i|-|m_j||-1}{2}\sin([|m_i|-|m_j|]\alpha_t) J_{||m_i|-|m_j||}(2\pi\beta|\tv|\kappa)\nonumber \\
&+& 2i\pi (-1)^\frac{||m_i|+|m_j||-1}{2}\sin([|m_i|+|m_j|]\alpha_t) J_{||m_i|+|m_j||}(2\pi\beta|\tv|\kappa)
\end{eqnarray}
\item \underline{$m_i>0$ and $m_j=0$:}
\begin{equation}
F(\kappa) =\sqrt{2} \int_0^{2\pi} \cos(|m_i|\alpha) \exp\left[2i\pi\beta|\tv|\kappa \cos(\alpha-\alpha_t)\right]{d}{\alpha}\nonumber
\end{equation}
\begin{enumerate}
\item if $m_i$ is even:
\begin{equation}
F(\kappa) = 2\pi\sqrt{2}(-1)^\frac{|m_i|}{2}\cos(|m_i|\alpha_t) J_{|m_i|}(2\pi\beta|\tv|\kappa)
\end{equation}
\item if $m_i$ is odd:
\end{enumerate}
\begin{equation}
F(\kappa) = 2i\pi\sqrt{2}(-1)^\frac{|m_i|-1}{2}\cos(|m_i|\alpha_t) J_{|m_i|}(2\pi\beta|\tv|\kappa)
\end{equation}
\item \underline{$m_i<0$ and $m_j=0$:}
\begin{equation}
F(\kappa) =\sqrt{2} \int_0^{2\pi} \sin(|m_i|\alpha) \exp\left[2i\pi\beta|\tv|\kappa \cos(\alpha-\alpha_t)\right]{d}{\alpha}\nonumber
\end{equation}
\begin{enumerate}
\item if $m_i$ is even:
\begin{equation}
F(\kappa) = 2\pi\sqrt{2}(-1)^\frac{|m_i|}{2}\sin(|m_i|\alpha_t) J_{|m_i|}(2\pi\beta|\tv|\kappa)
\end{equation}
\item if $m_i$ is odd:
\end{enumerate}
\begin{equation}
F(\kappa) = 2i\pi\sqrt{2}(-1)^\frac{|m_i|-1}{2}\sin(|m_i|\alpha_t) J_{|m_i|}(2\pi\beta|\tv|\kappa)
\end{equation}
\item \underline{$m_i=0$ and $m_j>0$:}
\begin{equation}
F(\kappa) =\sqrt{2} \int_0^{2\pi} \cos(|m_j|\alpha) \exp\left[2i\pi\beta|\tv|\kappa \cos(\alpha-\alpha_t)\right]{d}{\alpha}\nonumber
\end{equation}
\begin{enumerate}
\item if $m_j$ is even:
\begin{equation}
F(\kappa) = 2\pi\sqrt{2}(-1)^\frac{|m_j|}{2}\cos(|m_j|\alpha_t) J_{|m_j|}(2\pi\beta|\tv|\kappa)
\end{equation}
\item if $m_j$ is odd:
\end{enumerate}
\begin{equation}
F(\kappa) = 2i\pi\sqrt{2}(-1)^\frac{|m_j|-1}{2}\cos(|m_j|\alpha_t) J_{|m_j|}(2\pi\beta|\tv|\kappa)
\end{equation}
\item \underline{$m_i=0$ and $m_j<0$:}
\begin{equation}
F(\kappa) =\sqrt{2} \int_0^{2\pi} \sin(|m_j|\alpha) \exp\left[2i\pi\beta|\tv|\kappa \cos(\alpha-\alpha_t)\right]{d}{\alpha}\nonumber
\end{equation}
\begin{enumerate}
\item if $m_j$ is even:
\begin{equation}
F(\kappa) = 2\pi\sqrt{2}(-1)^\frac{|m_j|}{2}\sin(|m_j|\alpha_t) J_{|m_j|}(2\pi\beta|\tv|\kappa)
\end{equation}
\item if $m_j$ is odd:
\end{enumerate}
\begin{equation}
F(\kappa) = 2i\pi\sqrt{2}(-1)^\frac{|m_j|-1}{2}\sin(|m_j|\alpha_t) J_{|m_j|}(2\pi\beta|\tv|\kappa)
\end{equation}
\item \underline{$m_i=0$ and $m_j=0$:}
\begin{equation}
F(\kappa) = \int_0^{2\pi}\exp\left[2i\pi\beta|\tv|\kappa \cos(\alpha-\alpha_t)\right]{d}{\alpha} =  2\pi J_{0}(2\pi\beta|\tv|\kappa)
\end{equation}
\end{enumerate}
Putting all together the different combinations leads to the general expression of Eq. (\ref{eq_scalingtrans_final}).

\end{document}